\documentclass[structabstract,longauth]{aa}
\usepackage{txfonts}
\usepackage{times}
\usepackage{natbib}
\usepackage{graphicx}
\usepackage{color}
\usepackage{ulem}
\usepackage[switch, modulo]{lineno}
\newcommand{\RNum}[1]{\uppercase\expandafter{\romannumeral #1\relax}}

\begin{document}

\title{Search for Extended $\gamma$-ray Emission around AGN with
H.E.S.S. and \textit{Fermi}-LAT}
\author{H.E.S.S. Collaboration
\and A.~Abramowski \inst{1}
\and F.~Aharonian \inst{2,3,4}
\and F.~Ait Benkhali \inst{2}
\and A.G.~Akhperjanian \inst{5,4}
\and E.~Ang\"uner \inst{6}
\and G.~Anton \inst{7}
\and M.~Backes \inst{8}
\and S.~Balenderan \inst{9}
% ...
\email{stefan.ohm@le.ac.uk}}
\institute{
Universit\"at Hamburg, Institut f\"ur Experimentalphysik, Luruper Chaussee 149, D 22761 Hamburg, Germany \and
Max-Planck-Institut f\"ur Kernphysik, P.O. Box 103980, D 69029 Heidelberg, Germany \and
Dublin Institute for Advanced Studies, 31 Fitzwilliam Place, Dublin 2, Ireland \and
National Academy of Sciences of the Republic of Armenia, Yerevan  \and
Yerevan Physics Institute, 2 Alikhanian Brothers St., 375036 Yerevan, Armenia \and
Institut f\"ur Physik, Humboldt-Universit\"at zu Berlin, Newtonstr. 15, D 12489 Berlin, Germany \and
Universit\"at Erlangen-N\"urnberg, Physikalisches Institut, Erwin-Rommel-Str. 1, D 91058 Erlangen, Germany \and
University of Namibia, Department of Physics, Private Bag 13301, Windhoek, Namibia \and
University of Durham, Department of Physics, South Road, Durham DH1 3LE, U.K.
% ...
}

\date{}

\abstract {Very-high-energy (VHE; E$>$100\,GeV) $\gamma$-ray emission
  from blazars inevitably gives rise to electron-positron pair
  production through the interaction of these $\gamma$-rays with the
  Extragalactic Background Light (EBL). Depending on the magnetic
  fields in the proximity of the source, the cascade initiated from
  pair production can result in either an isotropic halo around an
  initially beamed source or a magnetically broadened cascade flux.}  {Both
  extended pair halo (PH) and magnetically broadened cascade (MBC) emission
  from regions surrounding the blazars 1ES~1101-232, 1ES~0229+200 and
  PKS~2155-304 were searched for, using VHE $\gamma$-ray data taken with
  the High Energy Stereoscopic System (H.E.S.S.), and high energy (HE;
  100\,MeV$<$E$<$100\,GeV) $\gamma$-ray data with the \textit{Fermi} Large
  Area Telescope (LAT).}  {By comparing the angular distributions of
  the reconstructed $\gamma$-ray events to the angular profiles calculated from 
  detailed theoretical models, the presence of PH and MBC was investigated.}
  {Upper limits on the extended emission around 1ES~1101-232,
  1ES~0229+200 and PKS~2155-304 are found to be at a level of few
  percent of the Crab nebula flux above 1\,TeV, depending on the
  assumed photon index of the cascade emission.  Assuming strong
  Extra-Galactic Magnetic Field (EGMF) values, $>10^{-12}\,{\rm G}$,
  this limits the production of pair halos developing from
  electromagnetic cascades.  For weaker magnetic fields, in which
  electromagnetic cascades would result in magnetically broadened
  cascades, EGMF strengths in the range $(0.3 - 3)$ $\times
  10^{-15}$\,G were excluded for PKS~2155-304 at the 99\% confidence
  level, under the assumption of a 1\,Mpc coherence length.}{}
 \titlerunning{Search for
  Extended Emission around AGN}
\maketitle

\newpage
\section{Introduction}
\label{Introduction}

About 50 Active Galactic Nuclei\footnote{See
  http://tevcat.uchicago.edu for an up-to-date list.}(AGN), with
redshifts ranging from 0.002 to 0.6, have so far been detected in
very-high-energy (VHE;\,E $>$ 100~GeV) $\gamma$-rays. Significant
emission beyond TeV energies has been measured for half of them.  The
spectra of such TeV-bright AGN with redshifts beyond $z\sim0.1$ are
significantly affected by the extragalactic background light (EBL)
\citep{nikishov:1962a, jelley:1966a, 1966PhRvL..16..252G}, with the
$\gamma$-rays from these sources interacting with the EBL and
generating electron-positron pairs.  The pairs produced, in turn, are
deflected by the Extra-Galactic Magnetic Field (EGMF) and cool by
interacting both with the EGMF via the synchrotron effect and with the
Cosmic Microwave Background (CMB) via the inverse Compton (IC)
effect. Thus, cascades can develop under certain conditions, with the
emerging high-energy photons being unique carriers of information
about both the EBL \citep{1993ApJ...415L..71S} and EGMF
\citep{2009PhRvD..80l3012N}.

Should the electron-positron pairs pass the bulk of their energy into
the background plasma through the growth of plasma instabilities
\citep{2012ApJ...752...22B}, a high-energy probe of the EGMF could be
invalidated. The growth rate of such instabilities, however, remains
unclear and is under debate \citep{2012ApJ...758..102S,
  2013ApJ...770...54M}. This work is conducted under the premise that
the IC cooling channel of the pairs dominates over any plasma cooling
effects.

\begin{table*}[]
\label{table:EGMF_regimes}
\caption{EGMF strength regimes for no cascade, pair halo and magnetically
  broadened cascade development. The effects of synchrotron losses for
  multi-TeV electrons in different EGMF strengths are
  summarised. ${\lambda_{\rm IC}}$ and $\tau_{\rm IC}$ represent the
  mean free path and cooling time for Inverse Compton interactions
  with the CMB, respectively.}
\begin{tabular}{lccc}
  \hline\hline
  regime number & \RNum{1} & \RNum{2} & \RNum{3}\\
  characterised by & synchrotron losses 
  & $2\pi r_g \ll \lambda_{\rm IC}$ & $2\pi r_{g} \gg c \tau_{\rm IC}$ \\
  \hline
  EGMF strength &B$ > 3\times 10^{-6}$G & $10^{-7}\mathrm{G}>$ B $> 10^{-12}$ G  & B $< 10^{-14}$G\\
  synchrotron losses & dominate over IC losses & negligible   & negligible  \\
  electromagnetic cascades & no cascade & pair halos & magnetically broadened cascades \\
  \hline\hline
\end{tabular}
\end{table*}

The strength of the EGMF has a major impact on the development of the
cascades. In order to explain its effects, three regimes of EGMF
strength are introduced in Table \ref{table:EGMF_regimes}. For strong
magnetic fields ($>10^{-7}$\,G, regime~\RNum{1}), synchrotron cooling
of pair-produced electrons becomes non-negligible, suppressing the
production of secondary $\gamma$-rays \citep{1978ApJ...225..318G}.
For such a scenario, the observed, $J_{\rm obs.}$, and intrinsic,
$J_{0}$, $\gamma$-ray fluxes are related as $J_{\rm obs.}(E)=J_{0}(E)
\exp[-{\tau_{\gamma \gamma}}(E,z)]$. Here, $\tau_{\gamma \gamma}(E,z)$
is the pair-production optical depth, which depends on the photon
energy $E$ and the redshift of the source $z$, as well as on the level
of the EBL flux $F(\lambda,z)$, where $\lambda$ is the EBL wavelength.

A weaker EGMF assumption removes the simple relation between the
observed and intrinsic energy spectra. For magnetic field strengths
between 10$^{-7}$\,G and 10$^{-12}$\,G (regime~\RNum{2}), the electron
pairs produced are isotropised and accumulate around the source,
eventually giving rise to a pair halo of secondary $\gamma$-rays
\citep{1994ApJ...423L...5A}. Since the isotropisation takes place on
much smaller scales than the cooling, the size of this pair halo
depends mainly upon the pair-production length, with very little
variation being introduced by the actual strength of the EGMF in the
above-mentioned range. The observed flux thus consists of both primary
and secondary high-energy $\gamma$-rays and its relation to the
intrinsic spectrum cannot be reduced to the simple effect of
absorption described by the optical depth (e.g.,
\citealt{2011A&A...529A.144T, 2011ApJ...731...51E}). Furthermore, the
level of secondary $\gamma$-rays emitted by the population of
accumulated pairs within the halo is able to provide a natural record
of the AGNs past activity \citep{1994ApJ...423L...5A}.

Unfortunately, due to the low $\gamma$-ray flux of the sources, and/or
a possible cutoff in the spectra below 10~TeV, combined with the
limited sensitivity of current generation $\gamma$-ray telescopes, the
detection of these halos in VHE $\gamma$-rays cannot be
guaranteed. Due to strong Doppler boosting, the apparent $\gamma$-ray
luminosities of AGN can significantly exceed the intrinsic source
luminosity \citep{1985ApJ...295..358L}. Furthermore, leptonic models
for many of the currently observed blazars do not require high photon
fluxes beyond 10\,TeV.

For even weaker magnetic field values (regime~\RNum{3},
B~$<~10^{-14}$\,G), no pair halo is formed, and the particle cascade
continues to propagate along the initial beam direction, broadening
the beam width. The angular size of this magnetically broadened
cascade (MBC) is dictated by the EGMF strength, and a measurement of
the broadened width can provide a strong constraint on the EGMF
value. Complementary to this probe, the combined spectra of the TeV
and GeV $\gamma$-ray emission observed from a blazar can also be used
as a probe of the intervening EGMF, as demonstrated in
\cite{2010Sci...328...73N}, \citet{2011A&A...529A.144T} and
\citet{2012arXiv1210.2802A}. Although generally the intergalactic
magnetic field is expected to have a much higher value, current
observations and cosmological concepts cannot exclude that in so
called ``voids'', with sizes as large as 100\,Mpc, the magnetic field
can be as small as $10^{-17}$\,G \citep{Miniati:2010ne,
  2013arXiv1303.7121D}. In the case of such weak ``void'' fields,
instead of persistent isotropic pair halo emission, the observer would
see direct cascade emission propagating almost rectilinearly over
cosmological distances.

As for the case of pair halos, the arriving flux in magnetically
  broadened cascades is also naturally expected to consist of a
  mixture of primary and secondary $\gamma$-rays. For flat or soft
intrinsic spectra (i.e., photon indices $\Gamma \gtrsim$~2), cascade
photons are expected to constitute a sub-dominant secondary component,
making measurements of the EBL imprint in blazar spectra possible
\citep{2007A&A...475L...9A, 2013A&A...550A...4H,
  2012A&A...542A..59M}. If, in contrast, the primary spectrum of
$\gamma$-rays is hard and extends beyond 10\,TeV, then, at lower
energies, the secondary radiation can even dominate over the primary
$\gamma$-ray component. Thus, in this case, the deformation of the
$\gamma$-ray spectrum due to absorption would be more complex than in
the case discussed above.

In a very small magnetic field, cascades initiated at very high
energies lead to the efficient transfer of energy back and forth
between the electron and $\gamma$-ray components, effectively reducing
the $\gamma$-ray opacity in the energy range of secondary particles
\citep{2002A&A...384..834A, 2011ApJ...731...51E}. As a result, the
observer is able to see $\gamma$-rays at energies for which
$\tau_{\gamma \gamma} \gg 1$.

On the other hand, due to deflections in the EGMF, the original
$\gamma$-ray beam is broadened, and even extremely small EGMF values
($\sim$$10^{-15}$~G) are expected to produce detectable extended
$\gamma$-ray emission. This radiation should be clearly distinguished
from that of pair halos. The origin of the extended emission in these
two cases is quite different, with pair halos producing extended
emission isotropically and MBCs producing extended emission only in
the jet direction.

The radiation from both pair halos and MBCs can be recognised by a
distinct variation in intensity with angular distance from the centre
of the blazar. This variation is expected to depend weakly on the
orientation and opening angle of the jet, and more on the total
luminosity and duty cycle of the source at energies $\geq$ 10\,TeV
\citep{1994ApJ...423L...5A}. To a first order approximation, the
radiation deflection angles remain small in comparison to the angular
size of the jet. Since the observer remains ``within the jet'', the
angles (relative to the blazar direction) from which the observer can
receive the magnetically broadened emission remain roughly independent
of the observer's exact position within the jet cone. This result,
however, only holds true if the observer is not too close to the edge
of the jet.

The preferred distance for the observation of both pair halos and MBCs
with the H.E.S.S. experiment is in the range of hundreds of
mega-parsecs to around one giga-parsec, i.e., in the range of
$\sim$0.1 to $\sim$0.24 in redshift. The far limit is set by the
reduction in flux with distance down to that only just sufficient for
detection. The near limit for pair halos results from the fact that
for sources too close, it becomes impossible to distinguish between
their halo photons and background radiation, as the halo would take up
the entire field of view of the observing instrument, i.e. 5$^{\circ}$
for H.E.S.S. For MBCs, similar near and far limits are found. In this
case, however, the near limit originates purely from a lack of cascade
luminosity: it becomes significant only for distances beyond several
pair production lengths.

A first search for pair halo emission was conducted by the HEGRA
collaboration \citep{2001A&A...366..746A} using Mkn~501 observations
(z = 0.033). This yielded an upper limit of $(5 - 10)$\% of the Crab
Nebula flux (at energies $\geq$ 1TeV) on angular scales of $0.5
^{\circ}$ to $1 ^{\circ}$ from the source. The MAGIC collaboration
performed a similar search for extended emission using Mkn~421 and
Mkn~501 \citep{2010A&A...524A..77A}. Upper limits on the extended
emission around Mkn~421 at a level of $<$ $5\%$ of the Crab Nebula
flux were obtained and a value of $<$ $4\%$ of the Crab Nebula flux
was achieved for Mkn~501, both above an energy threshold of
300\,GeV. These results were used to exclude EGMF strengths in the
range of few times 10$^{-15}$\,G. Since both Mkn~421 and Mkn~501 are
very nearby, the extension of the halo emission is expected to be
large. They are therefore no ideal candidates for this work.

More recently, a study was performed using data from the
$\textit{Fermi}$ Large Area Telescope (LAT)
\citep{2010ApJ...722L..39A}. Images from the 170 brightest AGN in the
11 month $\textit{Fermi}$ source catalogue were stacked
together. Evidence has been claimed for MBCs in the form of an excess
over the point-spread function with a significance of
3.5$\sigma$. However, \citet{2011A&A...526A..90N} showed that the
angular distribution of $\gamma$-rays around the stacked AGN sample is
consistent with the angular distribution of the $\gamma$-rays around
the Crab Nebula, (which is a point-like source for $\textit{Fermi}$)
indicating systematic problems with the LAT point spread function
(PSF).

In the latest publication on this topic \citep{2013ApJ...765...54A},
pair halo emission around AGN detected with \textit{Fermi}-LAT was
investigated with an updated PSF. A sample of 115 BL Lac-type AGN was
divided into high (z $>$ 0.5) and low redshift (z $<$ 0.5) blazars and
their stacked angular profiles were tested for disk and gaussian
shaped pair halo emission with extensions of 0.1$^{\circ}$,
0.5$^{\circ}$ and 1.0$^{\circ}$ by employing a Maximum Likelihood
Analysis in angular bins. No evidence for pair halo emission was found
in contrast to the results presented in \citet{2010ApJ...722L..39A},
and upper limits on the fraction of pair halo emission relative to the
source flux are given for three energy bins in the stacked
samples. Additionally, for 1ES~0229+200 and 1ES~0347-121, two BL Lac
objects that show $\gamma$-ray emission at TeV energies, upper limits
on the energy flux assuming different pair halo radii are given for
energies between 1 and 100\,GeV.

In this paper, a search for TeV $\gamma$-ray pair halos and
magnetically broadened cascades surrounding known VHE $\gamma$-ray
sources is presented. This study utilises both \textit{Fermi}-LAT and
H.E.S.S. data from three blazars. The three AGN selected,
1ES~1101-232, 1ES~0229+200, and PKS~2155-304, were observed between
2004 and 2009 with H.E.S.S. These AGN are in the preferable redshift
range and have emission extending into the multi-TeV energy domain,
thus making them ideal candidates for this study.

\section{Data Sets \& Analyses}
\label{Data_Analysis}

\subsection{H.E.S.S. Observations and Analysis Methods}

The H.E.S.S. experiment is located in the Khomas Highland of Namibia
($23 ^{\circ} 16^{\prime} 18^{\prime \prime}$S, $16 ^{\circ}
30^{\prime} 0^{\prime \prime}$E), 1835\,m above sea level
\citep{2004NewAR..48..331H}. From January 2004 to July 2012, it was
operated as a four telescope array (phase-I). The Imaging Atmospheric
Cherenkov Telescopes (IACT) from this phase are in a square formation
with a side length of 120\,m. They have an effective mirror area of
107\,m$^{2}$, detect cosmic $\gamma$-rays in the 100\,GeV to 100\,TeV
energy range and cover a field of view of $5^{\circ}$ in diameter. In
July 2012, a fifth telescope, placed in the middle of the original
square, started taking data (phase-II). With its 600\,m$^{2}$ mirror
area, H.E.S.S. will be sensitive to energies as low as several tens of
GeV.

For this analysis, only data from H.E.S.S. phase-I were used. To
improve the angular resolution, only observations made with all four
phase-I telescopes were included. Standard H.E.S.S. data quality
selection criteria \citep{2006A&A...457..899A} were applied to the
data set of each source. All data passing the selection were processed
using the standard H.E.S.S. calibration
\citep{2004APh....22..109A}. \textit{Standard cuts} \citep{Benbow2005}
were used for the event selection and the data was analysed with the
H.E.S.S. analysis package (HAP, version 10-06). The \textit{Reflected
  Region} method \citep{2006A&A...457..899A} was used to estimate the
$\gamma$-ray like background. Circular regions with a radius of
$\sqrt{0.22}^\circ$ around the sources were excluded from background
estimation in order to avoid a possible contamination by extended
emission from pair halos or MBCs.

The significance (in standard deviations, $\sigma$) of the observed
excess was calculated following \citet{1983ApJ...272..317L}. All upper
limits are derived following the method of
\citet{1998PhRvD..57.3873F}.

Using the stereoscopic array of four IACTs, the PSF is characterised
by a 68$\%$ containment radius of $\sim$0.1 degrees
\citep{2006A&A...457..899A}. The distribution of the squared angular
distance between the reconstructed shower position and the source
position ($\theta^{2}$) for a point-like source peaks at $\theta^{2} =
0$ and displays the PSF width. The PSF is calculated from Monte-Carlo
simulations, taking into account the observation conditions (e.g. the
zenith angle and the optical efficiency of the system) of each
observation as well as the photon index of the source.

\begin{table*}[]
\label{table:sourceInfo}
\caption{Summary of the H.E.S.S. analysis results for
  1ES~1101-232, 1ES~0229+200 and PKS~2155-304. The redshift,
  live-time, number of ON and OFF source events, $\gamma$-ray excess
  and significance ($\sigma$), mean zenith angle ($Z_{mean}$), mean
  offset ($\psi_{mean}$), the range of the Modified Julian Date (MJD)
  for the observations and the photon index $\Gamma$ for each source
  are reported.}
\begin{tabular}{lcccccccccccc}
\hline\hline
Source Name &Distance  & $T_{live}$ & $N_{ON}$ & $N_{OFF}$ & Excess & $\sigma$ & $Z_{mean}$ & $\psi_{mean}$ & MJD--50000 & $\Gamma$\\
 &({$z$})& (hours)  & & & & & (deg.) & (deg.) & (days) &  \\
\hline
 1ES~1101-232 & 0.186  & 62.9 & 79426 & 78636 & 790 & 10.8 & 22 & 0.6 & 3110 -- 4482 & 3.1 \\
 1ES~0229+200 & 0.140  & 72.3 & 39569 & 38752 & 817 & 6.6 & 45 & 0.56 & 3316 -- 5150 & 2.6 \\
 PKS~2155-304$_{\mathrm{low\, state}}$ & 0.117 & 164.5   & 200374 & 168685 & 31689 & 52.2 & 19 & 0.56 & 3199 -- 5042 & 3.4 \\
\hline
 PKS~2155-304$_{\mathrm{flare}}$	& 0.117 & 5.6  & 17440 & 6041 & 11399 & 78 & 21 & 0.56 & 3945 -- 3947& 3.4 \\
\hline\hline
\end{tabular}
\label{sec:data}
\end{table*}
 
Three VHE $\gamma$-ray sources, 1ES~1101-232, 1ES~0229+200 and
PKS~2155-304, have been chosen for this study due to their strong
emission in the $>$TeV energy range and their location in the suitable
redshift range. With $\sim$170~hours of good quality data,
PKS~2155-304 is a particularly well suited candidate for this
investigation. A summary of the results from the analyses can be found
in Table ~\ref{sec:data}. The results presented below have been
cross-checked with an independent analysis, the \textit{Model
  Analysis} (\citealt{2009APh....32..231D}), which yields consistent
results.

\begin{figure}
\includegraphics[width=100mm]{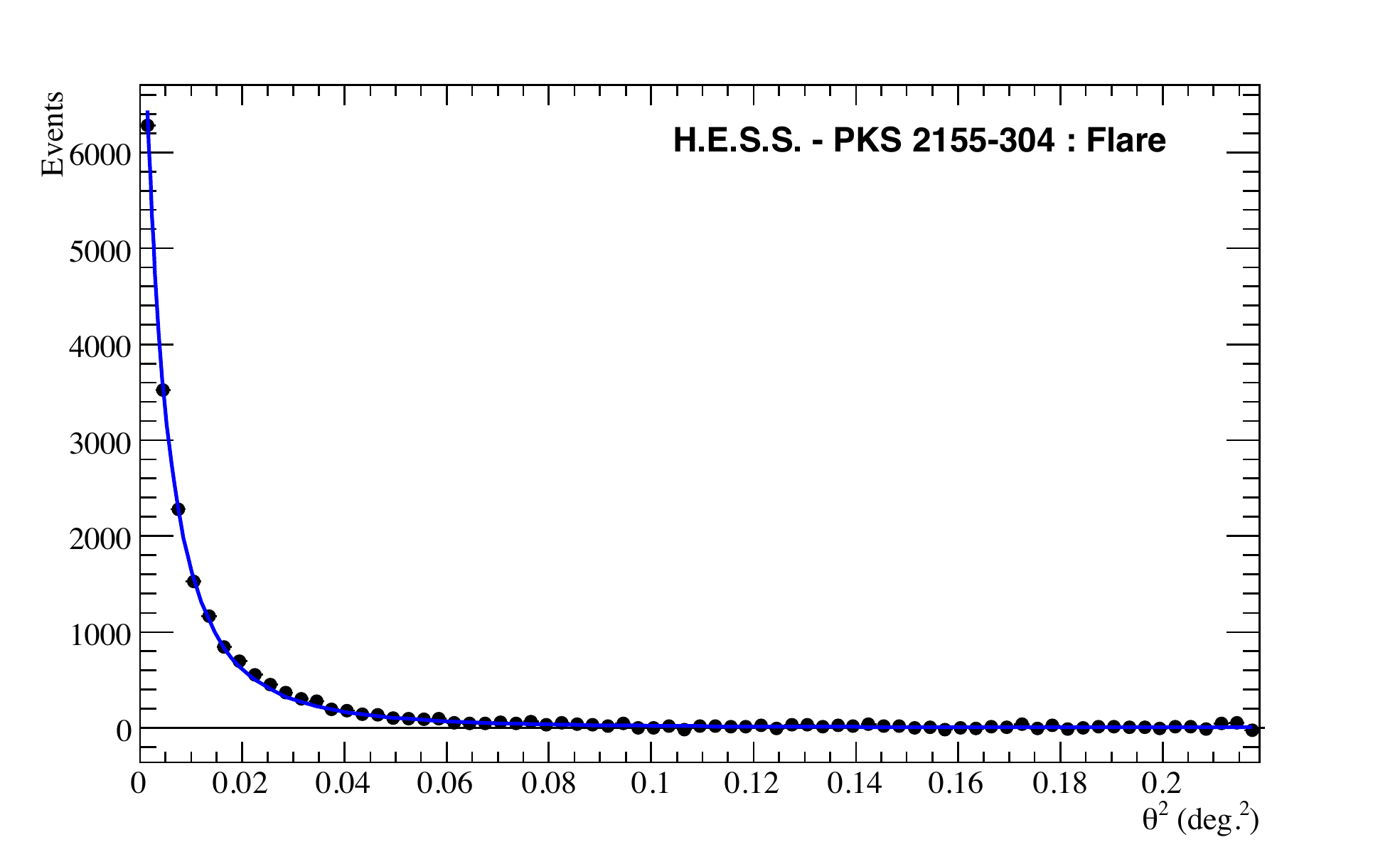}
\includegraphics[width=100mm]{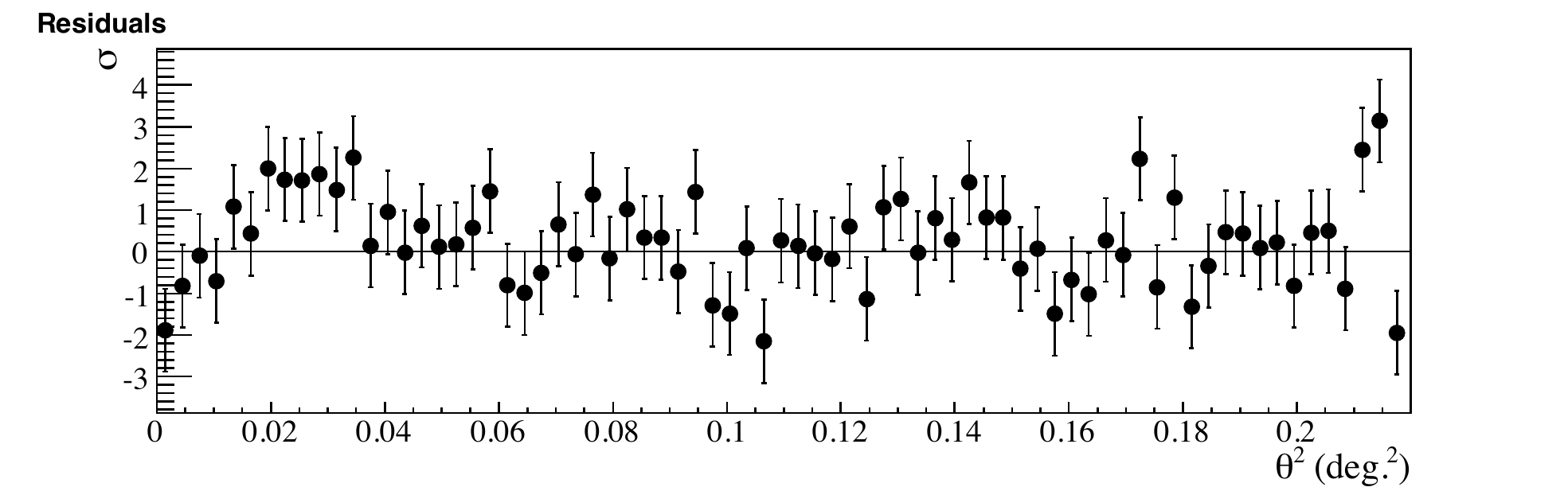}
\caption{Angular distribution of the PKS 2155-304 flare data set
  fitted with the H.E.S.S. point spread function (blue) from Monte
  Carlo simulations, resulting in a
  $\chi^{2}$/$\textit{n}_{\mathrm{d.o.f.}}$ = 91/72, with a
  P($\chi^{2}$) of 0.06. The fit residuals are shown in the lower
  panel.}
\label{sec:flare}
\end{figure}

\paragraph{\bf 1ES~1101-232}
The blazar 1ES~1101-232 was first discovered with H.E.S.S. in 2004 at
VHE $\gamma$-ray energies \citep{2007A&A...470..475A}. It resides in
an elliptical host galaxy at a redshift of \textit{z} = 0.186
\citep{1994ApJS...93..125F}. A total of $\sim$66~hours of good
quality data, taken between 2004 and 2008, have been analysed,
resulting in a detection significance exceeding 10$\sigma$.
\paragraph{\bf 1ES~0229+200}
This source was first observed by H.E.S.S. in late 2004 and detected
with a significance of 6.6$\sigma$ \citep{2007A&A...475L...9A}. This
high-frequency peaked BL Lac is hosted in a elliptical galaxy and is
located at a redshift of \textit{z} = 0.140
\citep{2005ApJ...631..762W}. A total of $\sim$80~hours of data taken
between 2004 and 2009 were used for this analysis. 1ES~0229+200 is a
prime source for such studies due to its hard intrinsic spectrum
reaching beyond 10\,TeV \citep{2007A&A...475L...9A,
  2012ApJ...747L..14V, 2010MNRAS.406L..70T, 2011ApJ...727L...4D}.
\paragraph{\bf PKS 2155-304}
Located at a redshift of $z$ = 0.117, PKS 2155-304 was first detected
with a statistical significance of 6.8$\sigma$ by the University of
Durham Mark 6 Telescope in 1999 \citep{1999ApJ...513..161C}. The
H.E.S.S. array detected this source in 2003 with high significance
($\sim$$45\sigma$) at energies larger than 160\,GeV
\citep{2005A&A...430..865A}. For this study, approximately 170~hours
of good quality data, taken between 2004 and 2009, have been analysed.
In 2006, this source underwent a giant outburst
\citep{2009A&A...502..749A}, with an integrated flux level ($>$ 200
GeV) about seven times that observed from the Crab Nebula. This value
is more than ten times the typical flux of PKS 2155-304 and the flux
varied on minute timescales. In the following, this exceptional
outburst is treated separately from the rest of the data, creating two
data sets: high state (i.e., the flare) and low state. Since the
pair-halo flux is not expected to vary on the time scales of the
primary emission, events in the flare data are mostly direct emission
from PKS~2155-304. Removing the flare from the main data set allows us
to focus on this source in a low state, where the contrast in flux
levels between primary and pair halo emission is smaller, facilitating
an easier detection. The data set for the low state amounts to
$\sim$165 hours, only including data of good quality. Focusing solely
on the exceptional flare from 2006, a data set corresponding to
$\sim$6~hours of observations was obtained during the nights of July
29th to 31st 2006. As described in \cite{2007ApJ...664L..71A}, the
short time scale ($\sim$200\,s) of the $\gamma$-ray flux variation
during the flare requires that the radius of the emission zone was $R
\delta^{-1}\leqslant 4.65 \times 10^{12}$\,cm in order to maintain
causality, $\delta$ being the Doppler factor. Considering the distance
of the source, the angular size of the emission region is therefore
$\leqslant 8 \times 10^{-9}$\,deg even with a minimal Doppler factor,
making it a point-like source for H.E.S.S.  The squared angular
distribution of the flare data set can be seen in
Fig. \ref{sec:flare}. It has been fitted with the H.E.S.S. PSF from
Monte-Carlo simulations resulting in a
$\chi^{2}$/$\textit{n}_{\mathrm{d.o.f.}}$ = 91/72, and a chance
probability P($\chi^{2}$) of 0.06. As can be seen from the residuals
in the lower panel of Fig. \ref{sec:flare}, the Monte-Carlo PSF
describes the data well, demonstrating that the flaring state is truly
consistent with being a point-like source for the instrument.

\par 

\subsection{\textit{Fermi}-LAT Analysis}
The $\textit{Fermi}$ Gamma-ray Space Telescope, launched in 2008,
observes the sky at energies between 20\,MeV and 300\,GeV
\citep{2009ApJ...697.1071A}. The $\textit{Fermi}$ data analysis
performed in this work used the LAT Science Tools package v9r23p1
(updated on 1st August~2011 to include the new PSFs) with the
P7SOURCE$\_$V6 post-launch instrument response function\footnote{See
  http://fermi.gsfc.nasa.gov/ssc/data/ for public \textit{Fermi} data
  and analysis software.}. The standard event selection for a source
well outside the galactic plane was applied. The analysis was
performed for SOURCE event class photons. The analysis was further
restricted to the energy range above 100\,MeV, where the uncertainties
in the effective area become smaller than 10\%.

The data used for this analysis corresponds to more than 4~years of
observations (4th August, 2008 - 1st March, 2013) for all three
sources. To produce the spectra and flux upper limits, binnedAnalysis
and UpperLimits Python modules were used, described in detail in the
$\textit{Fermi}$ data analysis threads. As is the standard procedure,
in order to take into account the broad $\textit{Fermi}$ PSF at low
energies, all sources from the Second \textit{Fermi}-LAT Catalog
(2FGL, \cite{FERMIcat}) within a 10-degree radius to the source
position were included. The energy range of 100\,MeV -- 300\,GeV was
split into logarithmically equal energy bins and in each bin a
spectral analysis was performed, fixing the power law index of each
source to be 2, and leaving the normalisation free. The normalisations
for Galactic and extragalactic backgrounds were also left free in each
energy bin. PKS~2155-304 and 1ES~1101-232 are detected in the dataset
above an energy threshold of 100\,MeV with significances of
$>$100$\sigma$ and 8.8$\sigma$, respectively. 1ES~0229+200 yields a TS
value of 31.7 which corresponds to a significance of about
5.6$\sigma$. The recent results on 1ES~0229+200 presented by
\citet{2012ApJ...747L..14V} are in agreement with the results
presented in this paper.

The spectra of the sources can be well fitted with a single power law
model with an index of $\Gamma=1.9\pm 0.2 $ for 1ES~1101-232,
$\Gamma=1.5\pm 0.3 $ for 1ES~0229+200 and $\Gamma=1.85\pm 0.02$ for
PKS~2155-304, with only statistical errors given. These spectral
indices are in good agreement with results from the 2FGL except for
1ES 0229+200, which was not listed in the catalogue.


\section{Pair Halo Constraints}
\label{PH_section}

\begin{figure*}[h!]
\centering
\begin{minipage}{0.45\textwidth}
\includegraphics[width=\textwidth]{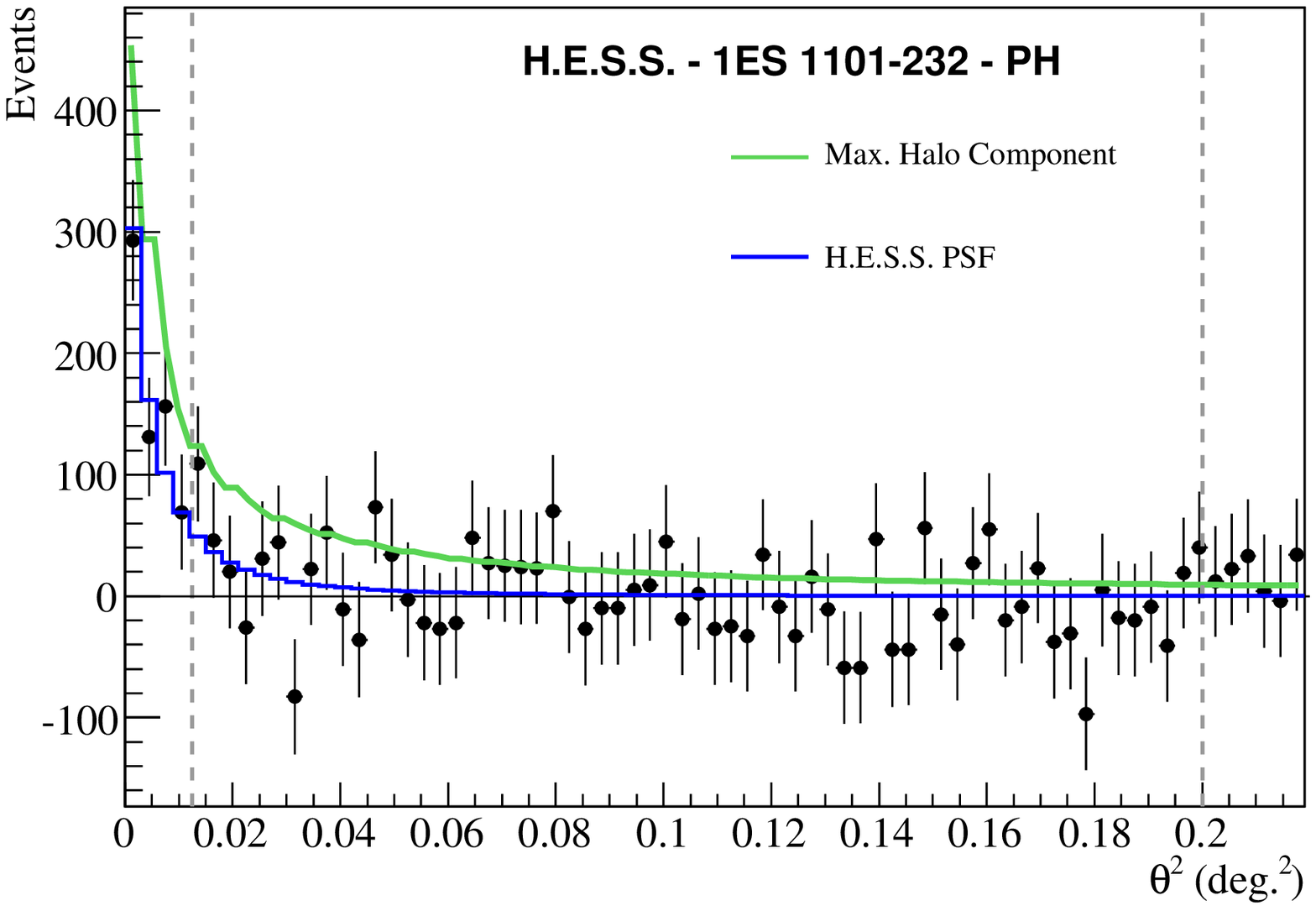}   
\includegraphics[width=\textwidth]{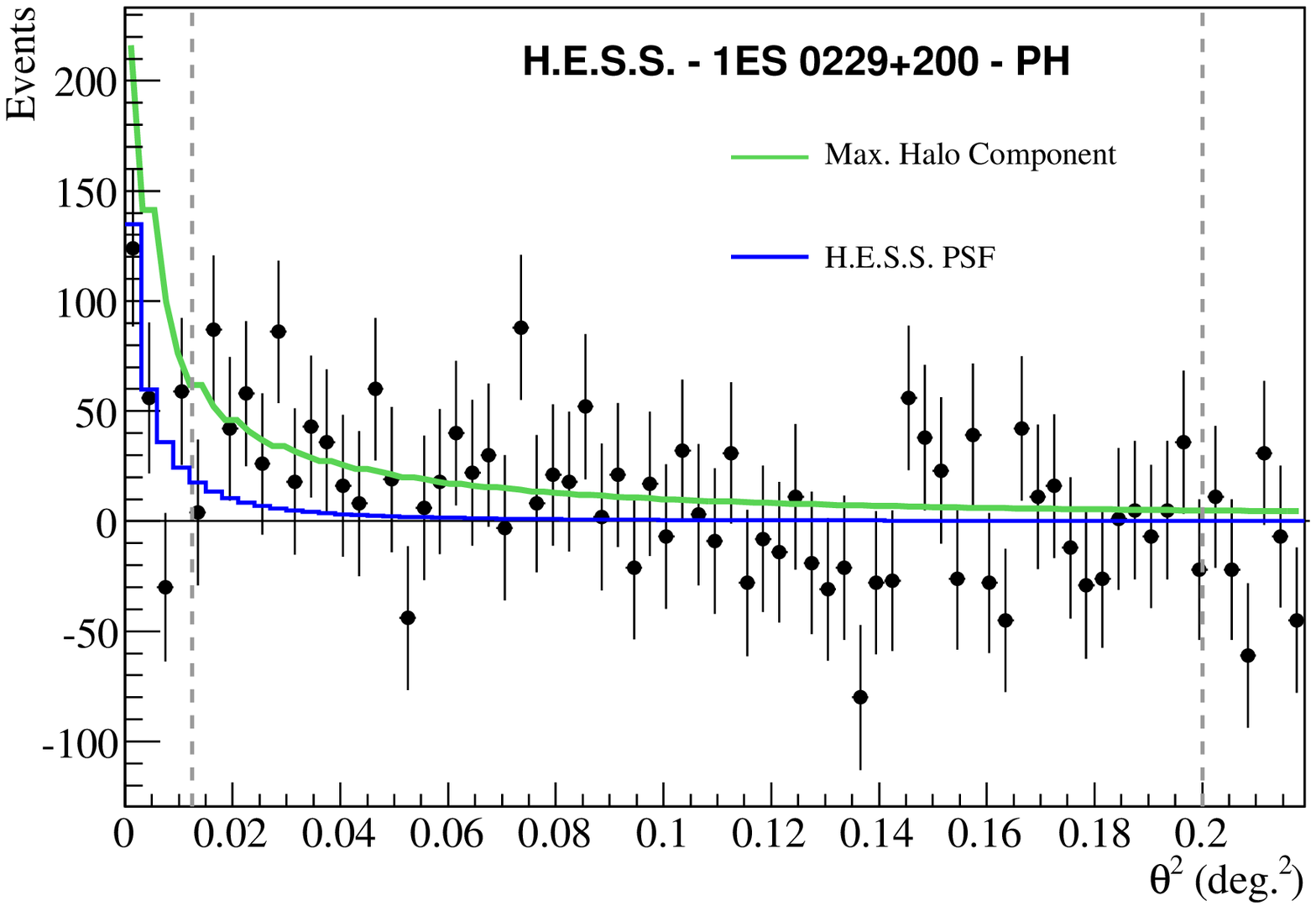}   
\includegraphics[width=\textwidth]{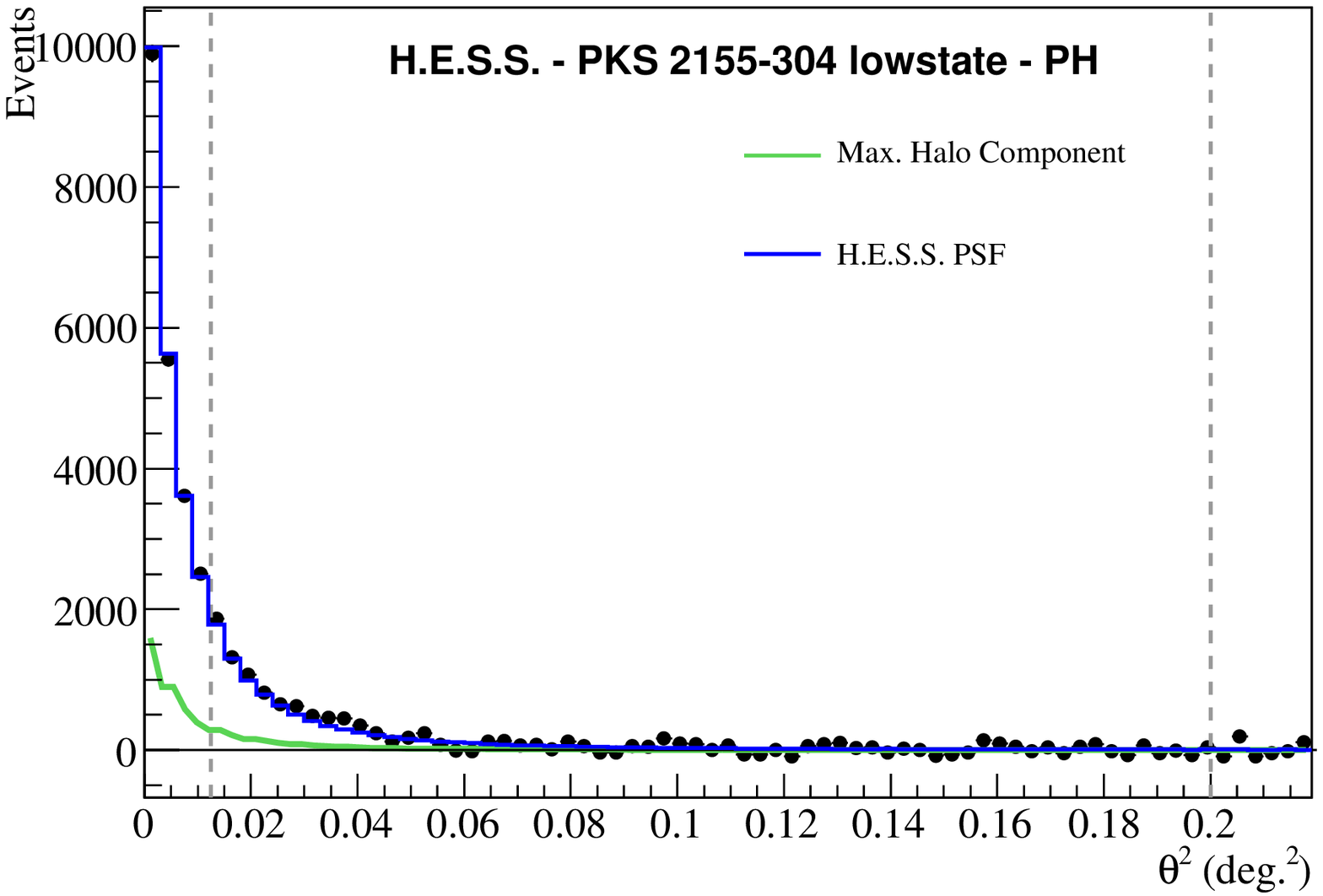}
\caption{Angular distribution of excess events of 1ES~1101-232 (top),
  1ES~0229+200 (middle) and the PKS~2155-304 low state (bottom). The
  blue line is the H.E.S.S. PSF and the green line is the maximum
  allowed halo component. The model independent limit on the pair halo
  excess is calculated between the vertical dashed lines at
  0.0125\,deg$^{2}$ and 0.02\,deg$^{2}$.}
\label{angle_all}
\end{minipage}\hfill
\begin{minipage}{0.45\textwidth}
\includegraphics[width=\textwidth]{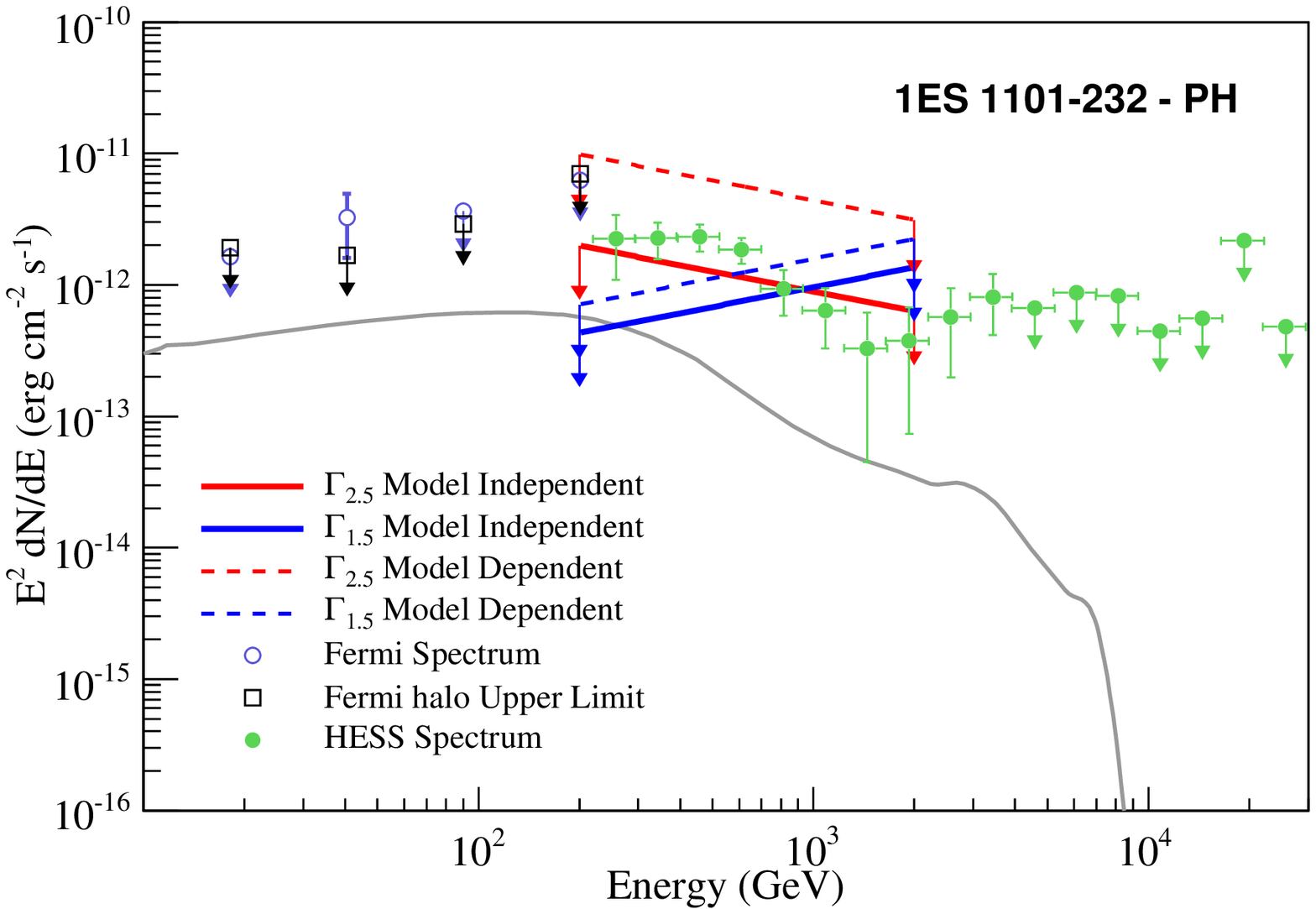}   
\includegraphics[width=\textwidth]{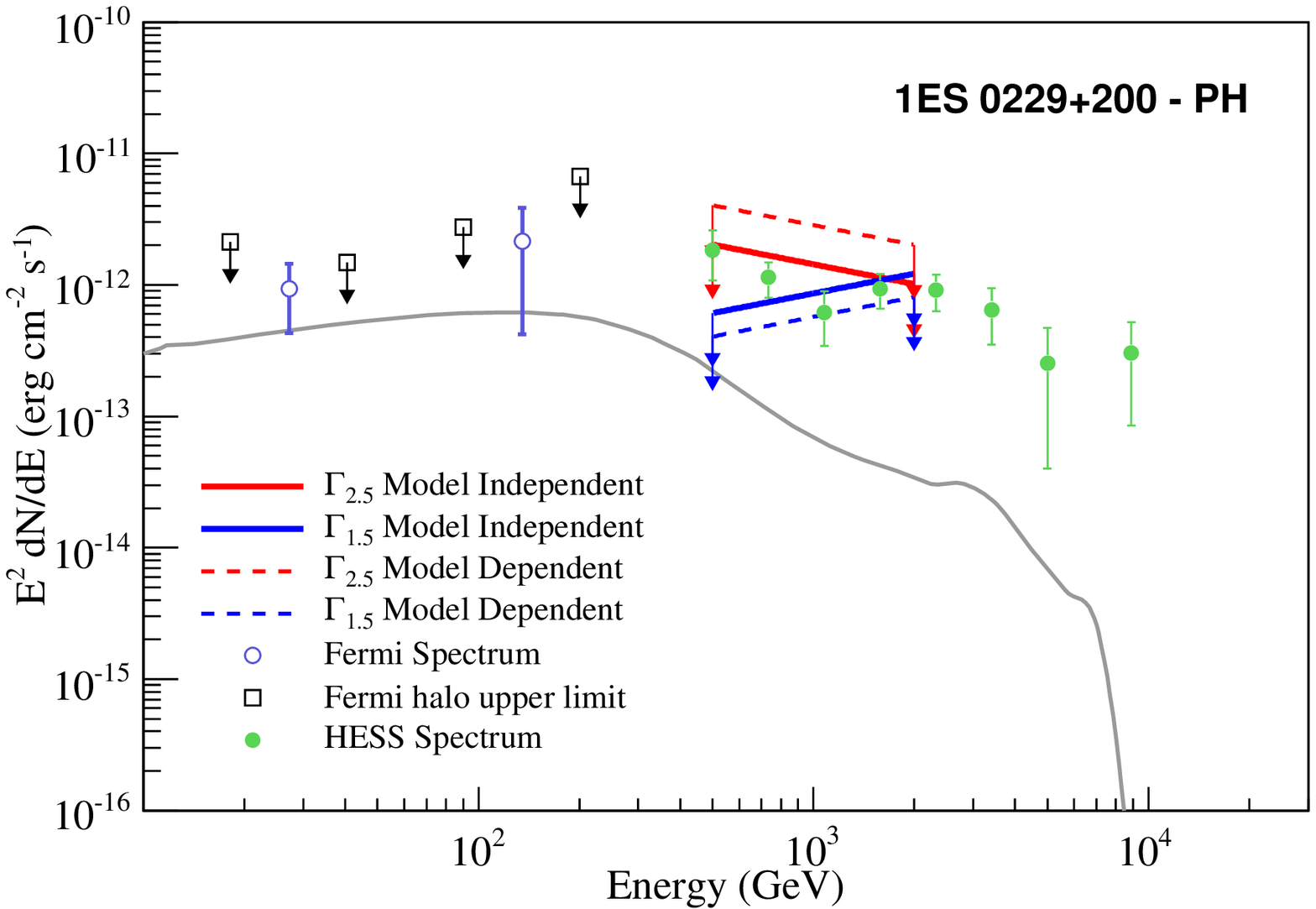}   
\includegraphics[width=\textwidth]{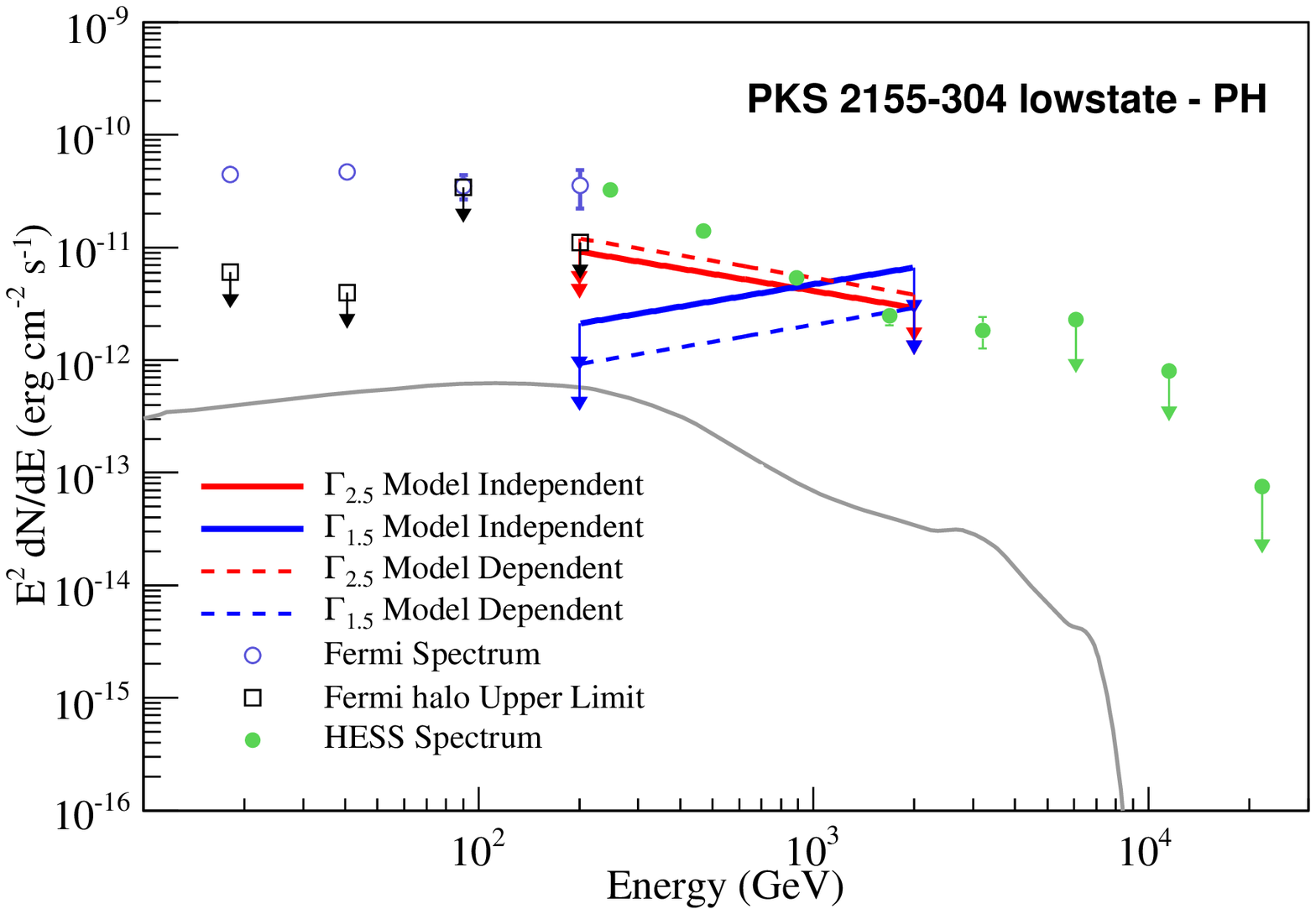}
\caption{Spectral energy distribution of 1ES~1101-232 (top),
  1ES~0229+200 (middle) and PKS~2155-304 low state data sets
  (bottom). The H.E.S.S. data (green circles) and the $\textit{Fermi}$
  data (empty circles) are shown. The upper limits on the flux
  contribution from a PH for the H.E.S.S. data are shown by blue and
  red arrows (dashed lines are model dependent and solid lines are
  model independent). The $\textit{Fermi}$ upper limits are shown as
  black squares. The grey line corresponds to the Halo Model taken
  from \citet{2009IJMPD..18..911E}.}
\label{sed_all}
\end{minipage}
\end{figure*}

Two separate techniques have been used to calculate pair halo (PH)
upper limits from H.E.S.S. data: a model dependent method and a model
independent method.  With each method, upper limits for two different
values of the photon index, 1.5 and 2.5, were calculated. These values
were chosen in order to illustrate the expected range of indices of
cascade emission at H.E.S.S. energies. A general model for the shape
of cascade spectra was developed in \cite{1988ApJ...335..786Z}. A more
recent model can be found in \cite{2009IJMPD..18..911E} and is
depicted as the grey curve in Fig.~\ref{sed_all}. Although predictions
at the high-energy end of the cascade strongly depend on the cutoff
energy of the injection spectrum, an index of $\sim$2 is expected in
the energy range just before the secondary flux drops rapidly. The
values 1.5 and 2.5 are representing a broader range of possibilities.
In addition, flux upper limits have been derived from
\textit{Fermi}-LAT data.

\paragraph{\bf Model Dependent Constraints}
In the publication by \citet{2009IJMPD..18..911E}, a study of the
formation of PHs was conducted. In particular, the authors
investigated the spectral and angular distributions of PHs in relation
to the redshift of the central source, the spectral shape of the
primary $\gamma$-rays and the flux of the EBL. In the results used
here from their study, the \citet{2001AIPC..558..463P} EBL model was
adopted.  In addition, the effects of the \citet{2008A&A...487..837F}
EBL model were investigated - these two models bound the present
uncertainties in the EBL in the relevant wavelength range to some
extent. Since the $(1-10)$\,$\mu$m EBL in the former model is
$\sim$40\% larger than in the latter, the upper limits on a possible
PH flux obtained with the \citet{2001AIPC..558..463P} EBL model are
more conservative. On the other hand, recent independent studies on
the EBL carried out by \citet{2013A&A...550A...4H} suggest an EBL
level between that motivated by the two EBL models considered.

For the \citet{2001AIPC..558..463P} EBL model, the differential
angular distribution of a PH at $\textit{z} \approx 0.13$ and
E$_{\gamma}$ $>$ 100\,GeV, which best suits our data, was taken from
Fig.~6 of \citet{2009IJMPD..18..911E} and is used here to derive
limits on a possible PH flux. The effect of the slight differences
between the assumed redshift in the model and the actual redshifts of
the analysed sources is smaller than the effect of different EBL
models and will therefore be neglected. The profile is based on
calculations employing mono-energetic primary $\gamma$-rays with an
energy E$_0$ = 100\,TeV. Provided the cutoff energy is sufficiently
high ($>$~5\,TeV), the differences in results for hard power-law and
mono-energetic injection scenarios are minor
\citep{2009A&A...502..749A, 2011A&A...526A..90N}. The resulting
angular distribution follows a profile of $\textit{ dN / d$\theta$
  $\propto$ $\theta^{-5/3}$}$. The angular distribution for the
\citet{2008A&A...487..837F} EBL model was generated by applying a
scaling relation. Though such a simple relation is not sufficient to
describe the effect of different EBL models on the angular shape of a
PH in general, it is appropriate for the energies and redshifts
discussed here (Eungwenichayapant \& Aharonian, private communication,
September 2013).

Using these spatial models, ``halo functions'' were created for the
measured $\theta^{2}$ distribution consisting of the PSF and the PH
angular profiles, convolved with the PSF: N($\theta^{2}$) =
N($\theta^{2})_{\mathrm{PSF}}$ + N($\theta^{2})_{\mathrm{PH}}$. The
PSF normalisation was left free and the number of excess events in the
PH model was increased until the fit had a probability $<$~0.05. With
this method, it was estimated how much of a halo component can be
added to the overall shape without contradicting observations at a
99$\%$ confidence level (C.L.).  In Fig.~\ref{angle_all}, the model
dependent analysis results for each of the three sources are shown,
under the assumption of the \citet{2001AIPC..558..463P} EBL model.
The green line in these figures represents the maximum possible halo
component allowed by the observational data. As can be seen in
Fig.~\ref{angle_all}a and b, due to low statistics, the total emission
for both 1ES~1101-232 and 1ES~0229+200 can be fitted with the halo
function. Therefore, the present angular profile data is unable to
significantly constrain a PH component. In contrast, a strong
constraint for a PH component of PKS~2155-304 could be derived:
relative to the central sources' flux, which is about five times
higher than the flux of 1ES~1101-232 and 1ES~0229+200, the upper limit
on a pair halo around PKS~2155-304 is the lowest. This is clearly
visible in Fig.~\ref{angle_all}c.  For the lower EBL fluxes predicted
by the \citet{2008A&A...487..837F} model, the upper limits in the
PKS~2155-304 case are even more constraining. Furthermore, although
the flux upper limits presented in Table \ref{Table_upper_limits} seem
high in comparison to the level of central point-like source fluxes,
one has to keep in mind that the limits derived here apply to a
comparatively large solid angle. The regions considered for the upper
limit calculation are 2.1~$\times 10^{-4}$\,sr (model dependent) or
1.99~$\times 10^{-4}$\,sr (model independent) while more than 75$\%$
of the flux from a point source as seen by H.E.S.S. are detected in a
region of 1.2~$\times 10^{-5}$\,sr, marked by the vertical line at
$\theta^{2} = 0.0125$~deg$^2$ in Fig.~\ref{angle_all}.

To determine the differential flux limit, the maximum number of halo
events was divided by the overall exposure, assuming a given photon
index. This method was repeated for two different values of the photon
index, 2.5 and 1.5. The resulting flux limits for both EBL models are
listed in Table~\ref{Table_upper_limits}. The upper limits on the PH
emission assuming the \citet{2001AIPC..558..463P} EBL model are shown
in Fig.~\ref{sed_all} together with the spectral energy distribution
(SED) of the sources. The H.E.S.S. spectral data are previously
published H.E.S.S. data taken from \citet{2007A&A...470..475A},
\citet{2007A&A...475L...9A} and \citet{2009ApJ...696L.150A},
respectively. Model dependent upper limits on the pair halo flux are
depicted as red lines for an assumed photon index of 2.5 and blue
lines for an assumed index of 1.5.

\paragraph{\bf Model Independent Constraints}
In the model independent approach, the residual emission after point
source subtraction was used to derive an upper limit on the PH
contribution. The expected contamination from the point-like source
was calculated by taking the integral of the PSF in the region
0.0125\,deg$^{2} < \theta^{2} <$ 0.2\,deg$^2$ (see vertical dashed
lines in Fig.~\ref{angle_all}), where the halo is expected to be most
dominant. The lower limit is chosen according to the standard
selection cut for point-like sources used by H.E.S.S. The Feldman
Cousins Confidence Intervals \citep {1998PhRvD..57.3873F} were used to
calculate the maximum halo excess at a 99\% C.L. Similarly to the
model dependent case, the differential limit was calculated by
dividing the maximum possible number of halo events by the overall
exposure, and the method was repeated for two different values of the
photon index (dashed blue and red lines in Fig.~\ref{sed_all}). In
several cases, conversely to that typically expected, the model
independent limits are more restricting than the model dependent
ones. This result is simply due to the poor statistics presently
available for the 1 ES objects.

\paragraph{\bf Constraints from \textit{Fermi}-LAT Data}
Since the pair halo is expected to be a diffuse source for
$\textit{Fermi}$, a spatial model ($\propto \theta^{-5/3}$) based on
theoretical estimations of the halo angular profile
\citep{2009IJMPD..18..911E} was used. The binned \textit{Fermi}
analysis was performed at energies 300 MeV -- 300 GeV for the models
with and without a halo component. In all considered cases, the models
with a halo have similar log-likelihood values to the models without
the halo contribution. Thus no significant indications for pair halo
emission are found. The upper limits on the fluxes at a 99\% C.L. were
calculated with the UpperLimits Python module of the \textit{Fermi}
software and are shown in Fig.~\ref{sed_all}.

\begin{table*}[]
\label{table:UpperLimits}
\caption{Pair halo flux upper limits for 1ES~1101-232, 1ES~0229+200
  and PKS~2155-304 at a 99\% C.L. All values are
  limits on the differential flux at 1~TeV given in units of
    $10^{-12}$\,TeV$^{-1}$\,cm$^{-2}$\,s$^{-1}$.}
\begin{center}
\begin{tabular}{lcc|cc|cc}
  \hline\hline
  & \multicolumn{4}{c}{Model Dependent} & \multicolumn{2}{c}{Model Independent}\\\hline
  Source Name & \multicolumn{2}{c|}{Franceschini EBL} & \multicolumn{2}{c|}{Primack EBL} & & \\

  & $\Gamma = 1.5$ & $\Gamma = 2.5$ & $\Gamma = 1.5$ & $\Gamma = 2.5$ & $\Gamma = 1.5$ & $\Gamma = 2.5$\\
\hline

1ES~1101-232 & 2.3 & 2.1 & 2.1 & 2.0 & 0.6 & 0.6 \\
1ES~0229+200 & 1.2 & 2.0 & 0.8 & 1.4  & 0.5 & 0.9 \\
PKS~2155-304$_{\mathrm{low\,state}}$ & 1.3 & 1.1 & 2.3 & 2.0 & 2.9 & 2.6 \\    
\hline\hline
\end{tabular}
\end{center}
\label{Table_upper_limits}
\end{table*}

\section{Magnetically Broadened Cascade Constraints}
\label{BBC_section}

In this section a model dependent approach was applied in order to
investigate whether evidence for a magnetically broadened cascade is
found in the angular event distribution of blazar fluxes observed with
H.E.S.S. A 3D Monte-Carlo description of magnetically broadened
cascades developed in \citet{2011A&A...529A.144T} was utilised here, in
order to determine the expected angular profile of this emission for
different EGMF strengths. For these calculations, both the
\citet{2008A&A...487..837F} and the \citet{2001AIPC..558..463P} EBL
model were used. Using this description, the range of EGMF values
excluded by the present H.E.S.S. results was investigated. A method
similar to the model dependent approach described in
Section~\ref{PH_section} was used to obtain these constraints: A
spatial MBC model function N($\theta^{2}$) =
N($\theta^{2})_{\mathrm{PSF}}$ + N($\theta^{2})_{\mathrm{MBC}}$ was
created, N($\theta^{2})_{\mathrm{MBC}}$ being the MBC model from
simulations convolved with the H.E.S.S. PSF. In the same manner as for
the model dependent PH limits, the PSF normalisation was left free and
the number of MBC events was increased until contradicting the
observational results at a 99\% C.L. The ratio of maximum allowed MBC
events was then compared to the ratios predicted by the Monte-Carlo
simulations for different magnetic field strengths. In the
simulations, photon indices of $1.9$, $1.5$, and $1.9$ for
1ES~1101-232, 1ES~0229+200, and PKS~2155-304, respectively, were
motivated from the \textit{Fermi} analysis of their GeV spectra. The
spectra of all three of the blazars used in this study are consistent
with a power-law spectrum with a cutoff at multi-TeV
energies. Therefore, for each of the sources an injection spectrum of
the form $dN/dE \propto E^{-\Gamma}e^{-E/E_{\rm max}}$ with a cutoff
$E_{\rm max}=10$\,TeV was adopted to ensure that a sufficient amount
of the cascade component lies in the H.E.S.S. energy range (see
\citealt{2009IJMPD..18..911E}).

\begin{figure*}[h!]
\centering
\begin{minipage}{0.45\textwidth}
\includegraphics[width=\textwidth]{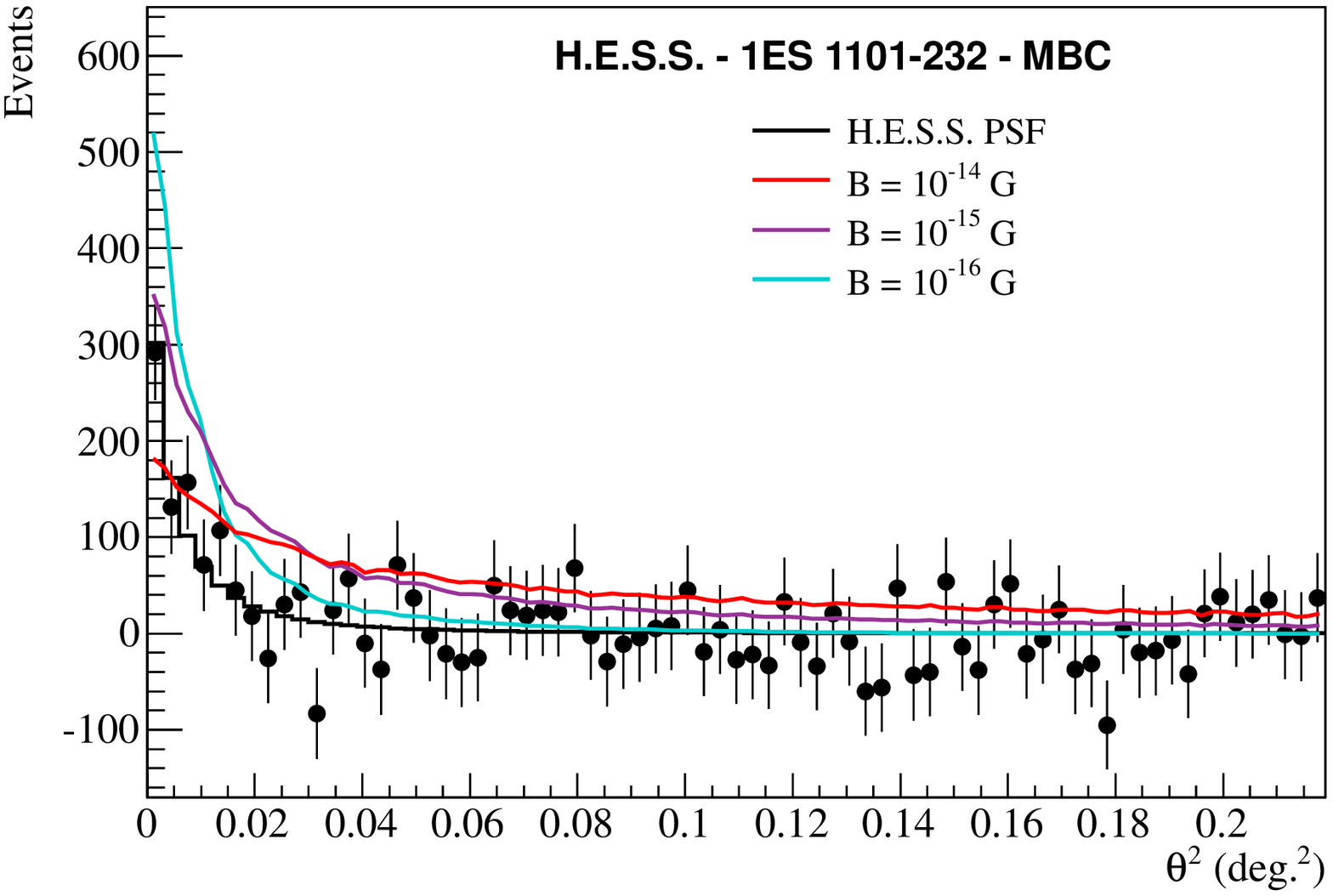}
\includegraphics[width=\textwidth]{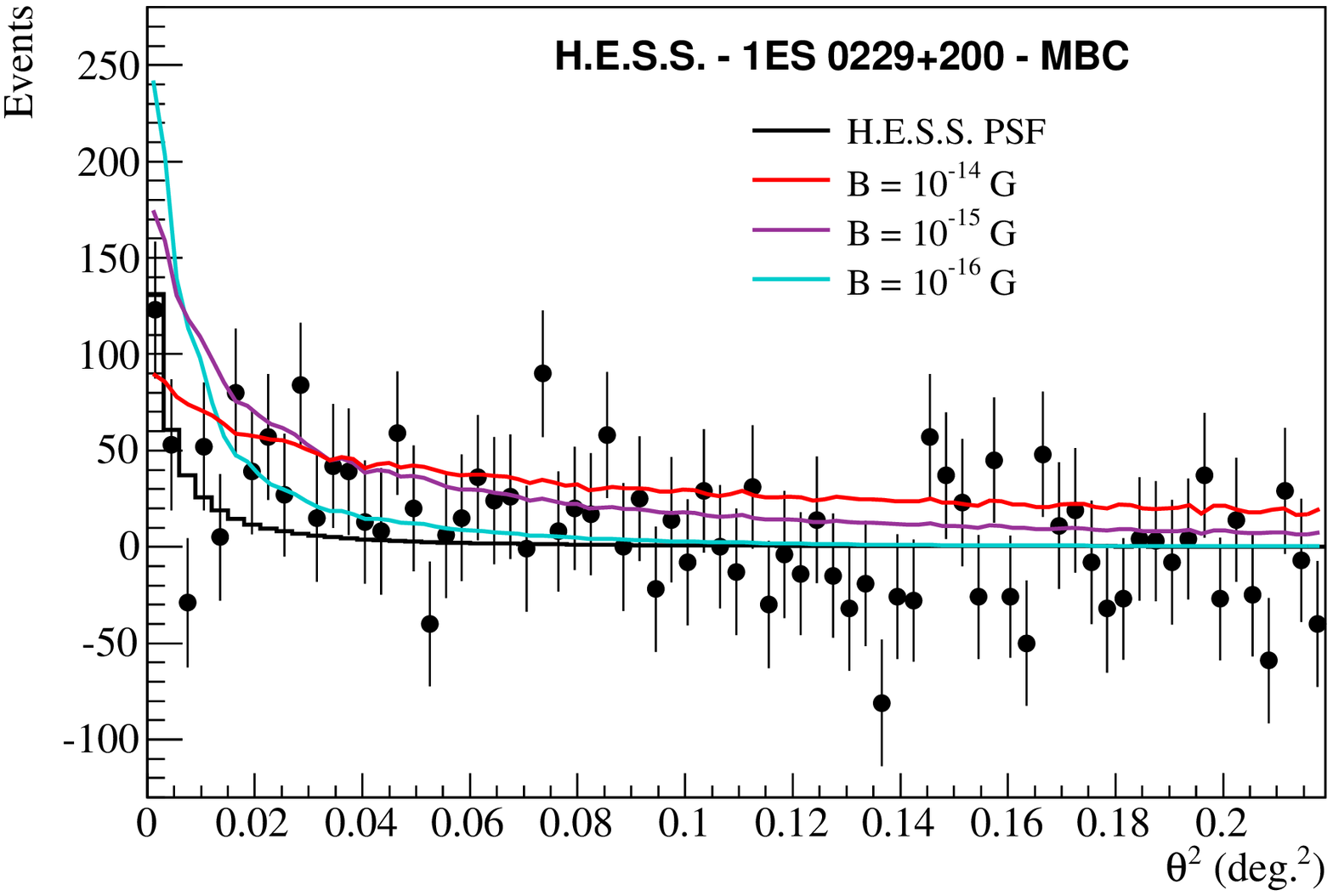}
\includegraphics[width=\textwidth]{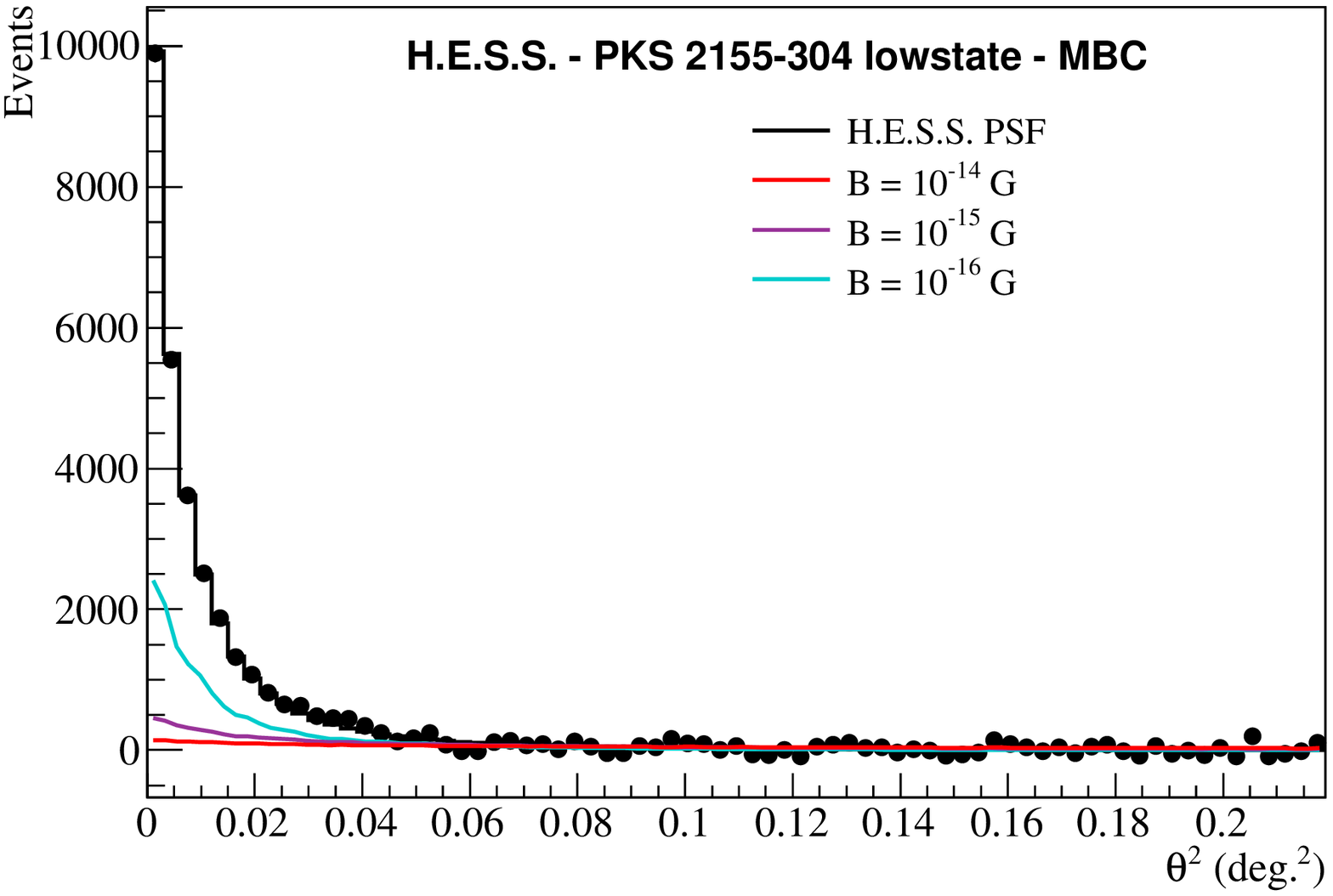}
\caption{Angular distribution of excess events of 1ES~1101-232 (top),
  1ES~0229+200 (middle) and the PKS~2155-304 low state (bottom). The
  H.E.S.S. data (black points) is plotted against the angular
  distribution of the magnetically broadened cascade model for varying
  magnetic field strengths.  The red, violet and cyan lines correspond
  to the maximum cascade flux for magnetic field strengths of
  10$^{-14}$, 10$^{-15}$ and 10$^{-16}$ G, simulated under the
  assumption of the \cite{2008A&A...487..837F} EBL model. }
\label{sec:angular_profiles}
\end{minipage}\hfill
\begin{minipage}{0.45\textwidth}
\includegraphics[width=\textwidth]{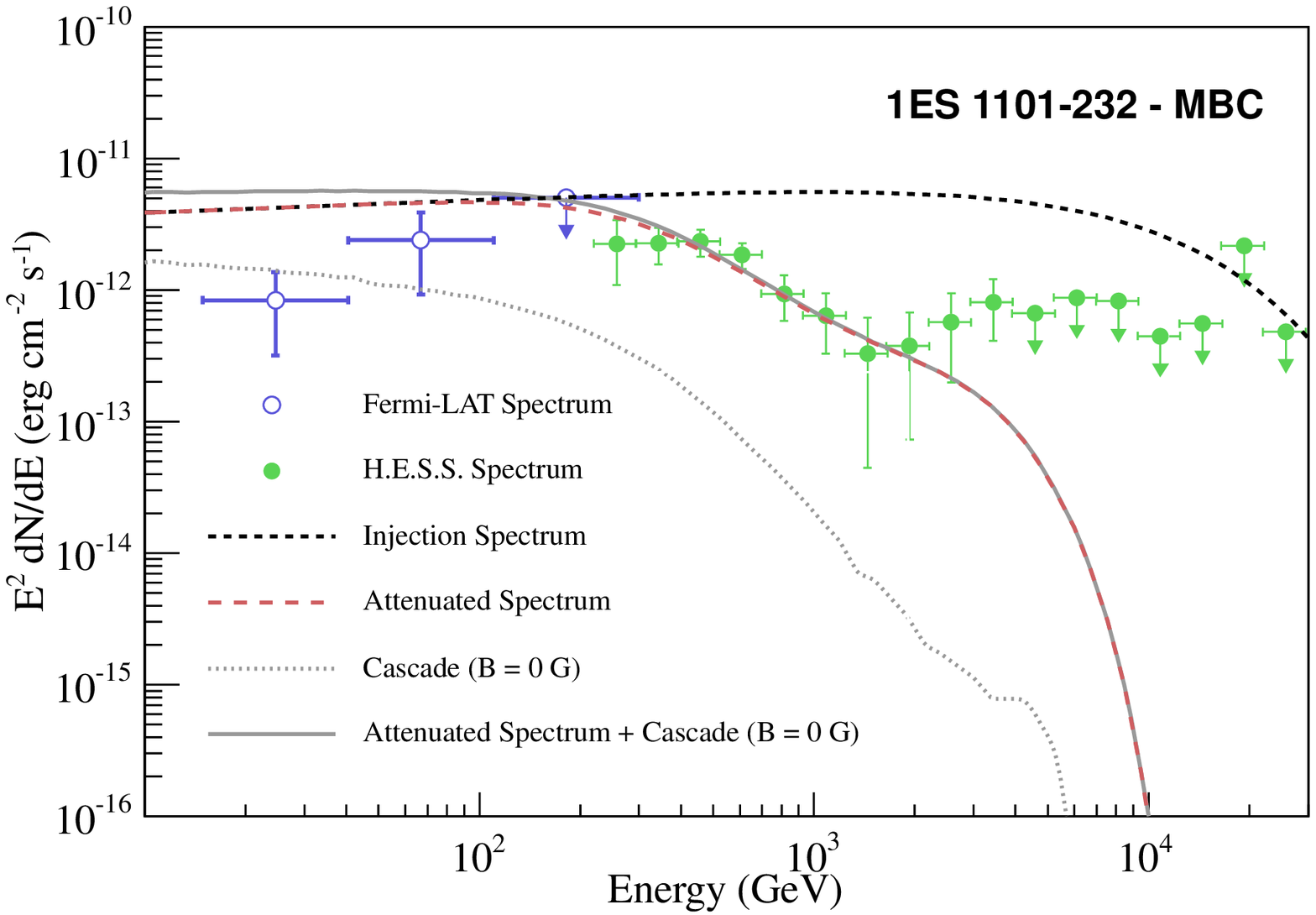}
\includegraphics[width=\textwidth]{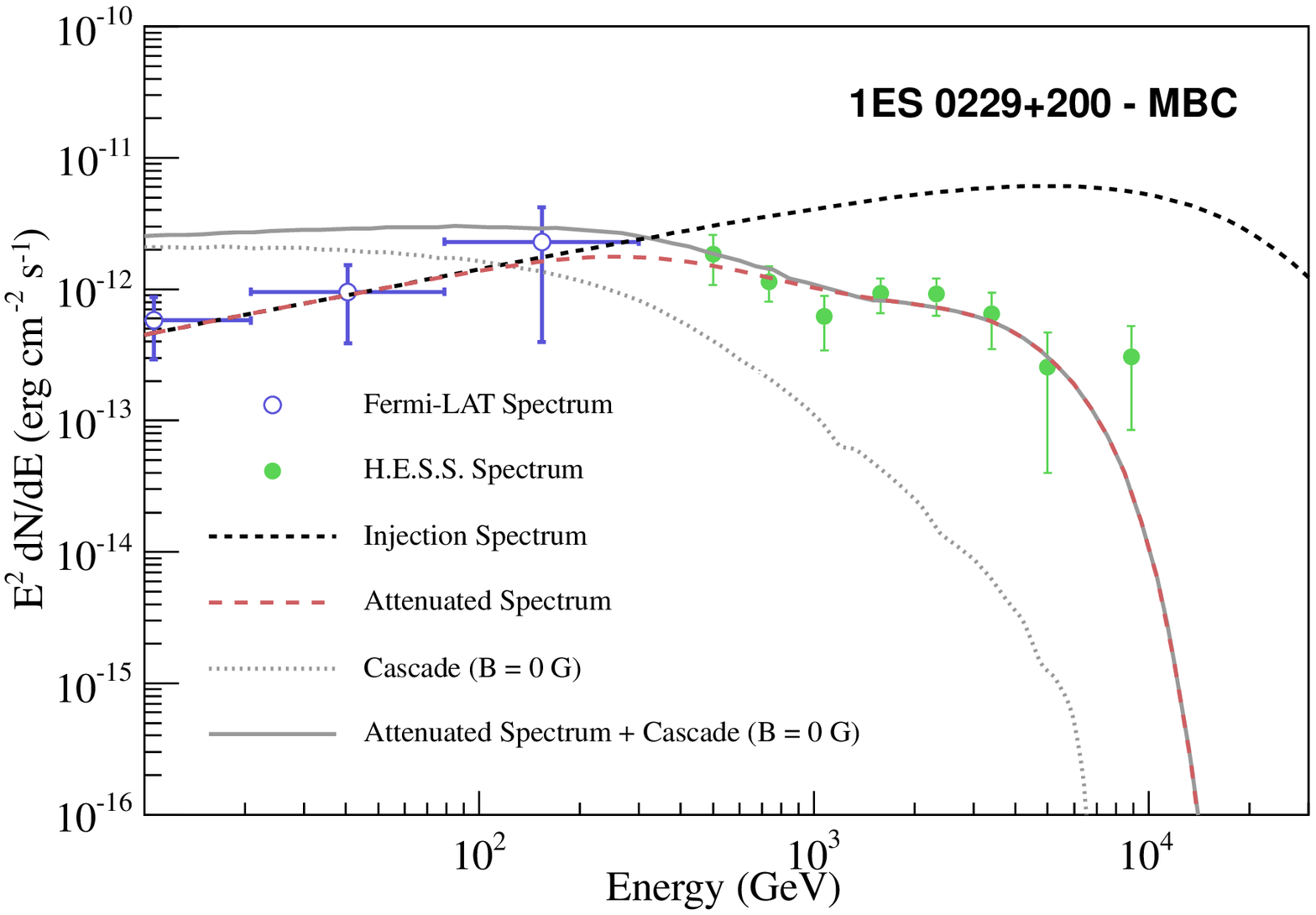}
\includegraphics[width=\textwidth]{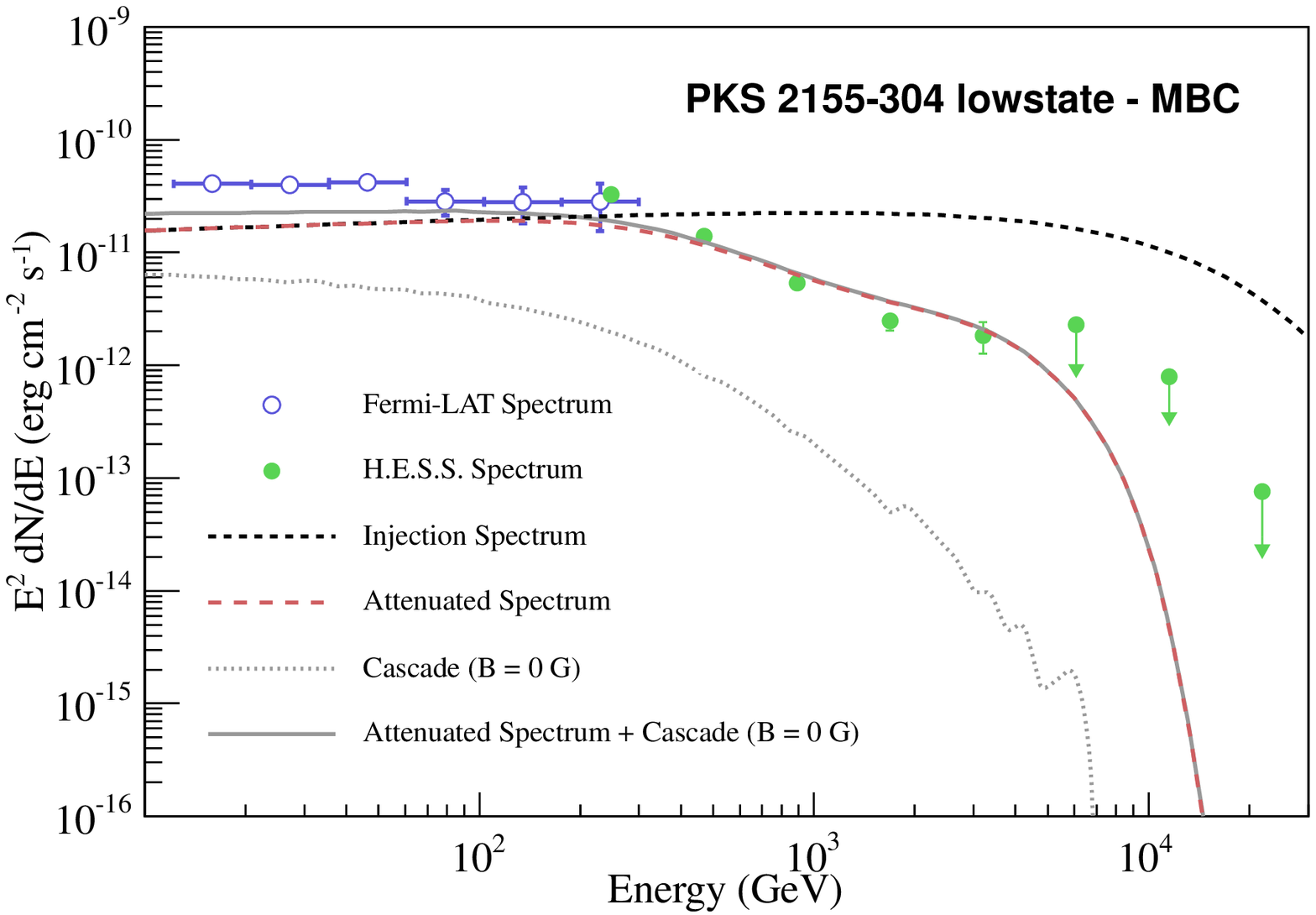}
\caption{The 1ES~1101-232 (top), 1ES~0229+200 (middle) and
  PKS~2155-304 (bottom) spectral energy distributions ($\Gamma$ = 1.9,
  1.5 and 1.9 respectively), including $\textit{Fermi}$ data (blue
  empty circles) as well as the H.E.S.S. results (green solid
  circles). The dotted grey line shows the expected cascade SED
  assuming the EGMF strength is $0$~G, and the solid grey line shows
  this component added to the attenuated direct emission SED (dashed
  red line).}
\label{sec:SEDs}
\end{minipage}
\end{figure*}

For the magnetically broadened cascade scenario, both the observed SED
and angular spread of the arriving flux depend significantly on the
EGMF. The angular spreading effect is seen explicitly in
Fig.~\ref{sec:angular_profiles}, for which the effect of
$10^{-14}$\,G, $10^{-15}$\,G, and $10^{-16}$\,G EGMF values are
considered. A 1\,Mpc coherence length is adopted as a fiducial value,
although larger values were discussed recently
\citep{2013arXiv1303.7121D}. Essentially, the effect of the coherence
length can be neglected if it is larger than the cooling length of the
multi-TeV cascade electrons of relevance here. In contrast, the choice
of the EBL model plays an important role. Again, the
\cite{2001AIPC..558..463P} EBL model is expected to result in more
conservative bounds on the maximum cascade contribution since it is
about 40\% higher than the \cite{2008A&A...487..837F} EBL model at
the wavelengths of interest here.

\begin{figure}[h]
\includegraphics[width=0.5\textwidth]{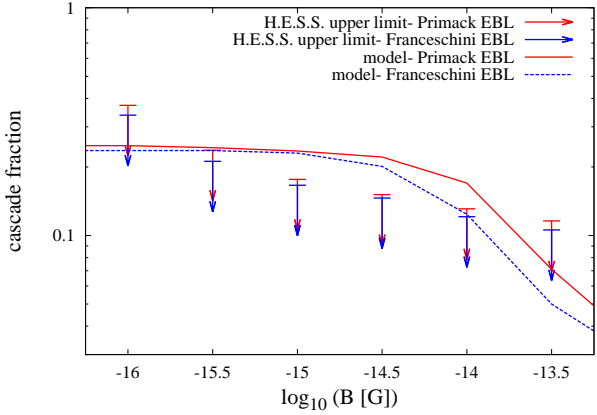}
\caption{EGMF constraints on PKS~2155-304. The blue dashed line
  depicts the expected fraction of MBC events in the VHE data
  depending on the EGMF strength, assuming the
  \cite{2008A&A...487..837F} EBL model. Blue arrows are the maximum
  fractions of MBC events for that EBL model not contradicting the
  angular profile data of PKS~2155-304 at a 99$\%$ C.L. The expected
  cascade fraction and the corresponding upper limit from
  H.E.S.S. data under the assumption of the \cite{2001AIPC..558..463P}
  EBL model are depicted in red.}
\label{Bfield_constraint}
\end{figure}

In Fig.~\ref{sec:angular_profiles}, the angular profiles of the MBCs
resulting from calculations with the
\textbf{\citet{2008A&A...487..837F}} model are shown. Though the
comparably low statistics for both 1ES~1101-232 and 1ES~0229+200 limit
any constraint from their measured angular profiles, a strong
constraint is provided by the angular profile of PKS~2155-304. For
this object, a mild cascade contribution was found to be expected in
the arriving VHE photon flux. As can be seen in
Fig.~\ref{Bfield_constraint}, for PKS~2155-304 the maximum ratio of
MBC events in the H.E.S.S. data is in conflict with the expected ratio
of cascade photons introduced by field strengths of
$\sim$$10^{-15}$\,G or a factor of a few stronger.  Assuming the
  \citet{2001AIPC..558..463P} EBL model, the range of excluded EGMF
  strengths is $(0.3 - 10)\times 10^{-15}$\,G. On the other hand, the
  \citet{2008A&A...487..837F} EBL model is the conservative choice
  when excluding EGMF strengths. Since it predicts a much
  lower cascade fraction for B = 10$^{-14}$G, such a magnetic field
  strength regime can not be ruled out when assuming this EBL model.
For stronger fields the cascade contribution's fraction to the overall
arriving flux, relative to that of the direct emission component,
reduces significantly due to isotropisation. Consequently, the
subsequent angular spreading for higher EGMF values becomes
indistinguishable from the H.E.S.S. PSF. Thus, for EGMF values such as
those present in the PH scenario discussed in
Section~\ref{PH_section}, the angular profiles can be significantly
smaller than those found for the case of a $10^{-15}$\,G EGMF
value. This strong EGMF suppression effect explains why the above
derived 99\% C.L. on the EGMF value constrains only a decade 
of EGMF range. In addition, all bounds depend on whether 
the intrinsic cutoff energy is high enough. For the two EBL models 
considered, \citet{2001AIPC..558..463P} and \cite{2008A&A...487..837F}, 
a minimum cutoff above 3\,TeV is required such that some constraint be
obtainable. For a larger cutoff energy than the value adopted in this
study, the range of excluded EGMF would be a few times larger.

\section{Discussion \& Conclusions}
\label{Discussion_Conclusion}

The search for a pair-halo component in the H.E.S.S. and
\textit{Fermi}-LAT data from regions surrounding the VHE $\gamma$-ray
sources 1ES~1101-232, 1ES~0229+200 and PKS~2155-304 shows no
indication for the presence of such emission. From our analysis, flux
upper limits on the extended VHE $\gamma$-ray emission from the three
sources analysed have been found to be at the level of a few percent
of the Crab Nebula flux. For example, the model independent upper
limits on the pair halo flux for an assumed photon index of 2.5 are
$<$~2\%, $<$~3\% and $<$~8\% of the integrated Crab Nebula flux above
1\,TeV\footnote{$(2.26 \pm 0.03) \times 10^{-11} cm^{-1} s^{-1}$, see
  \cite{2006A&A...457..899A}} for 1ES~1101-23, 1ES~0229+200 and
PKS~2155-304, respectively, adopting the \cite{2001AIPC..558..463P}
EBL model. Also with the analyses of \textit{Fermi}-LAT data, no
significant pair halo emission was detected and energy-binned flux
upper limits for a $\theta^{-5/3}$ profile were calculated. Though
these limits are comparable to previously obtained values by other
instruments for other blazars, the detailed angular modelling from
recent theoretical work on the topic, adopted by this study, marks a
significant improvement on previous limits. While the most
constraining upper limit values in \cite{2001A&A...366..746A},
\cite{2010A&A...524A..77A} and \cite{2013ApJ...765...54A} were derived
by varying the angular size of the extended emission model, the
analysis at hand gives all limits with a physically motivated fixed
size. However, with the method presented here, upper limits would
become more constraining the less similar the expected extended
emission is to the PSF. The constraints obtained from this pair halo
analysis can be used to set limits on the $\gamma$-ray output from
these AGN over the past $\sim$10$^{5}$ years. If any of these AGN had
been more active in the past, more pairs would have been subsequently
produced. Consequently, for sufficiently strong EGMF values
($>10^{-12}$\,G), increased activity in the past would strengthen the
constraint on the extended emission component. Since the EGMF strength
required for the pair halo scenario leads to the isotropisation of the
cascade emission, the observed luminosity of this secondary component
may be significantly reduced compared to the apparent luminosity of
the primary beamed component. A lack of detection of the secondary
component, therefore, is unable to place constraints on the EGMF
strength.

The limits of the PH $\gamma$-ray energy flux for the three blazars
may be converted into limits on the accumulated electron energy
density in the surrounding regions. As an example case, 1ES~0229+200
is considered, whose energy flux at 0.5\,TeV is
$\sim$$10^{-12}$\,erg\,cm$^{-2}$~s$^{-1}$. Assuming that the
corresponding photons result from a pair-halo cascade with strong EGMF
($>10^{-12}$\,G), the parent $\sim$$15$\,TeV electrons and positrons
will be both born into and isotropised within a region $\sim$$10$\,Mpc
from the blazar. For this strong field case an upper limit on the TeV
$\gamma$-ray luminosity from these regions is $\sim$$4\times
10^{42}$\,erg\,s$^{-1}$. Since the electron IC cooling time on the CMB
is $t^{e}_{\rm cool}(15\,{\rm TeV})\approx$ 10$^{5}$\,yr, a limit on
the total energy content of the parent electrons is
$\sim$$10^{55}$\,erg.

A search for MBC emission in the arriving flux from the three blazars
was also carried out. The datasets for both 1ES~1101-232 and
1ES~0229+200 were found not to be statistically constraining at
present. However, a constraint was found to be obtainable using the
PKS~2155-304 observational results. From H.E.S.S. observations of the
angular profile for PKS~2155-304, EGMF values were excluded for the
range $(0.3 - 3)\times 10^{-15}$\,G (for a coherence length of
$1$\,Mpc), at the 99\% C.L. This range is excluded for both EBL models
adopted here, the \cite{2001AIPC..558..463P} as well as the
\cite{2008A&A...487..837F} model. For a coherence length scale
$\lambda_{B}$ shorter than the cascade electron cooling lengths, the
lower EGMF limit scales as $\lambda_{B}^{-1/2}$, as demonstrated in
\cite{2013A&A...554A..31N}. Conversely, for $\lambda_{B}$ longer than
these cooling lengths, the constraint is independent of $\lambda_{B}$.
As shown in Fig.~6, stronger magnetic fields than the upper limit
result in the cascade component dropping below the direct emission
contribution, reducing the overall angular width below the H.E.S.S.
resolution limits.

Furthermore, our bound on the EGMF is compatible with the analytic
estimates put forward in \citet{2010A&A...524A..77A}, although the
analysis presented here is the most robust to date due to the
theoretical modelling that has been employed.

Interestingly, the success proven by this method demonstrates its
complementarity as an EGMF probe in light of the multi-wavelength SED
method employed in previous studies (\citealt{2010Sci...328...73N},
\citealt{2011ApJ...727L...4D}, \citealt{2011MNRAS.414.3566T} and
\citealt{2011A&A...529A.144T}).  These studies probed EGMF values for
which no notable angular broadening would be expected. Instead, the
effect of the EGMF was to introduce energy dependent time-delays on
the arriving cascade.  Ensuring that the source variability timescale
sits at a level compatible with that currently observed, i.e. the
sources are steady on $3$\,yr timescales, placed a constraint on the
EGMF at a level of $>10^{-17}$\,G \citep{2011A&A...529A.144T,
  2011ApJ...733L..21D}. In contrast to this time delay SED method, our
angular profile investigations are insensitive to the source
variability timescale. In this way, the constraints provided by the
angular profile studies of blazars offer a complementary new probe
into the EGMF, allowing field strengths with values $> 10^{-15}$\,G to
be investigated.

The future prospects for observing both extended halo emission and
MBCs are promising. In the near future, H.E.S.S. phase-II offers great
potential with its ability to detect $\gamma$-rays in the energy band
between \textit{Fermi}-LAT and H.E.S.S. phase-I. In the longer term,
the Cherenkov Telescope Array (CTA; see
e.g. \citealt{2013APh....43....3C}), with a larger array size, a wider
field of view, improved angular resolution along with greater
sensitivity will allow for a deeper probing of these elusive
phenomena.

\scriptsize{\textit{Acknowledgements}. The support of the Namibian
  authorities and of the University of Namibia in facilitating the
  construction and operation of H.E.S.S. is gratefully acknowledged,
  as is the support by the German Ministry for Education and Research
  (BMBF), the Max Planck Society, the German Research Foundation
  (DFG), the French Ministry for Research, the CNRS-IN2P3 and the
  Astroparticle Interdisciplinary Programme of the CNRS, the
  U.K. Science and Technology Facilities Council (STFC), the IPNP of
  the Charles University, the Czech Science Foundation, the Polish
  Ministry of Science and Higher Education, the South African
  Department of Science and Technology and National Research
  Foundation, and by the University of Namibia. We appreciate the
  excellent work of the technical support staff in Berlin, Durham,
  Hamburg, Heidelberg, Palaiseau, Paris, Saclay, and in Namibia in the
  construction and operation of the equipment. The authors are grateful to the referee who helped to considerably improve the quality of the paper.}


\bibliography{aa_references}



\end{document}